\documentclass[twocolumn,showpacs,preprintnumbers]{revtex4}

\usepackage{graphicx}
\usepackage{dcolumn}
\usepackage{bm}
\usepackage{amssymb}
\usepackage{epsfig}

\def\beq{\begin{equation}}
\def\eeq{\end{equation}}
\def\bea{\begin{array}}
\def\eea{\end{array}}
\def\be{\begin{equation}}
\def\ee{\end{equation}}
\def\ba{\begin{eqnarray}}
\def\ea{\end{eqnarray}}

\def\to{\rightarrow}

\def\f{\frac}
\def\[{\left[}
\def\]{\right]}
\def\({\left(}
\def\){\right)}

\def\gn{\tilde {\rm n}}

\def\to{\rightarrow}

\def\sq2{\sqrt{2}}

\def\End{\end{document}}

\begin{document}

\title{ FCNC production of same sign Top quark pairs \\
at the LHC. }  
\author{%
{ F. Larios} and { F. Pe$\gn$u$\gn$uri}
}
\affiliation{%
\vspace*{2mm} 
Departamento de F\'{\i}sica Aplicada, 
CINVESTAV-M\'erida, AP 73 Cordemex, 97310
M\'erida, Yucat\'an, M\'exico
}

\begin{abstract}
\hspace*{-0.35cm} 
We study the possibility of same sign top quark pair
production at the LHC (and the VLHC) as a direct probe of
FCNC processes.  Besides the SM neutral Z boson, two other
neutral bosons are considered, a top-Higgs type scalar and a
Z' boson that appear in Topcolor assisted Technicolor models.
We find that the FCNC couplings tqV (q=u,c ; V=H,Z,Z')
may produce an interesting signal of same sign top quark
pairs that could be observed at the LHC (VLHC).
\pacs{\,12.60.-i,\,12.15.-y,\,11.15.Ex 
\hfill   ~~ [ November 5, 2003 ] }

\end{abstract}

\maketitle

\setcounter{footnote}{0}
\renewcommand{\thefootnote}{\arabic{footnote}}


\section{Introduction}

With an integrated luminosity of about 100 fb$^{-1}$ the CERN LHC
is expected to produce several tens of millions of $t\bar t$
pairs for each of the detectors, ATLAS and CMS per
year\cite{konisberg}. 
Such a high rate of production will allow the LHC to look for
new physics effects involving the top quark.  In particular,
Flavor Changing Neutral Current (FCNC) effects are a well
known way to look for physics beyond the Standard Model (SM).
The standard way to measure FCNC couplings is
by searching for rare top quark decays like $t\to q\gamma, qZ$
($q=u,c$)\cite{raretopdec}.  Also, the FCNC couplings $Vtq$, with
$V=\gamma,\, Z,\, H$ and gluon can be probed via the associated
$tZ$, $t\gamma$ and $tH$ production at the
LHC\cite{multilepton,branco}, or at lepton colliders\cite{eetop}.

\begin{figure}
\includegraphics[width=6.1cm,height=3cm]{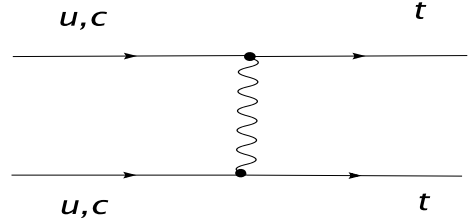}
\vspace*{0mm}
\caption{There are 2 Feynman diagrams, t-chanel and u-channel
for same sign tt production process.
The t-channel diagram is shown.
}
\label{diagram}
\end{figure}

The appearance of FCNC in the quark sector occurs at the one
loop level in the SM and is very suppresed by the GIM mechanism.
However, some extensions of the SM can generate these couplings
at tree level,
like Topcolor assisted Technicolor\cite{hill,theories}.
At the LHC one can think of a process in which two incoming
up type quarks exchange a neutral boson and then change to
a pair of (same sign) top quarks\cite{slabospitsky},
see Fig.~(\ref{diagram}).
As expected from the fact that the anomalous coupling
appears twice in the Feynman diagram, the rates predicted
can become negligible for not very small coupling values.
In this sense, the usual production process that involves only
one anomalous coupling is usually a better probe.  However,
we believe that the production of same sign top quark pairs 
is an interesting way to confirm the existence of a FCNC
with the top quark, and it should be considered in future
runnings of the LHC.  Apart from the standard backgrounds
associated with a top quark pair there is no other SM process
that can hide this signal.  The SM process $pp\to bbW^+W^+$ is
negligible.  Unfortunately, the only way to separate the $tt$
from the $t\bar t$ signal is by identification of same charges
in the dilepton mode -a mode with very low efficiency. 
However, the size of the FCNC couplings could be large and as
we shall see, the LHC could find many pairs of tops even during
the first few years of running.  Another interesting feature of
this signal is that higher center-of-mass energies of the
colliding beams would boost the production rates, which makes
this process a very good possibility at a future VLHC.

Below, we will discuss separately this process via three
intermediary bosons for the FCNC: the SM Z boson, a Z' boson
and a scalar boson that couples strongly with the top quark.
Then we will make a comment on the detection mode required
to separate the $tt$ signal from the $t\bar t$ signal.

\vspace*{-3mm}
\section{FCNC for the top quark.}
%

We will consider the two possibilities for the Flavor
transition of the top quark with the two lighter up and
charm quarks.
We parameterize the FCNC couplings of the top quark with
the following Lagrangian:
\ba
\L &=& \bar t ( y^R_{tq} P_R +  y^L_{tq} P_L ) q\; H
\nonumber \\ 
&+&  \; \bar t \,
\gamma_\mu \,(a^L_{tq} P_L + a^R_{tq} P_R)\,
q \, Z^\mu \nonumber \\
&+& g_2 \bar t \gamma_\mu \,(B^L_{tq} P_L +
B^R_{tq} P_R)\, q \, {Z'}^\mu  
\ea
where, $q=u,c$.
Similar parameterizations have appeared in the literature;
for instance, the couplings $a^{L,R}_{tq}$ are defined as
$a^{L,R}_{tq}\,\equiv\,\f{g_W}{2c_W} X^{L,R}_{tq}$ in
Ref.~\cite{multilepton, slabospitsky}, and
$y^{L,R}_{tq}\,\equiv\,\f{g_W}{2} g^{L,R}_{tq}$ with
$c_{a,v}=1/{\sq2}$ in Ref.~\cite{branco}.  In general, there
can be dimension 5 (and higher) operators that generate
transition magnetic dipole couplings $tq\gamma$, $tqZ$ and
$tq$-gluon\cite{multilepton,tcgluon, slabospitsky}.
In this work we are only considering dimension 4 interactions.

\subsection{FCNC with the Z boson.}

There are three top quark FCNC couplings that involve existing
gauge bosons $\gamma$, $Z$ and the gluon and that are
currently considered at the Tevatron Run II\cite{wwagner}.  Here,
we want to take the case of the Z boson.  Similar results will
hold for the other two bosons, as we shall see later.
At LEP all the FCNC couplings of the Z boson with the fermions
have been strongly constrained for all the charged leptons and
the light quarks.  However, the top quark cannot be a decay
product of Z, and the only direct constraint of a
$Ztq$~($q=u,c$) coupling comes from the Tevatron\cite{abe}:
\ba
\sqrt{|X^{R}_{tq}|^2 + |X^{L}_{tq}|^2} \leq 0.8 \, ,
\ea
which translates to $|a_{tq}|\leq 0.3$ for the couplings
defined here.  The weakness of
this bound is also evident from the associated upper bound
on the branching ratio $Br(t\to cZ) \leq 33\%$\cite{abe}.
However, Run II of the Tevatron is expected to reach a level
of $1.5\%$, which would be translated in a much stronger
bound on $X^R_{tq} \leq 0.04$\cite{wwagner}.
Actually, it has been shown that this coupling can be
tested down to values of order $10^{-2}$ at the LHC
via the associated production of a top and a Z or $\gamma$
\cite{multilepton}.
On the other hand, even the present weak bound can
be made stronger by considering more specific models.  In
Ref.~\cite{extraquarks} an extension of the SM with extra
quark singlets or doublets was considered; it was found that
present stringent constraints on the diagonal couplings
imply an $X_{tq} \leq 0.2$ ($a_{tq} \leq 0.07$) bound that
is one fourth the size of the direct limit.

\begin{figure}
\includegraphics[width=8.1cm,height=6cm]{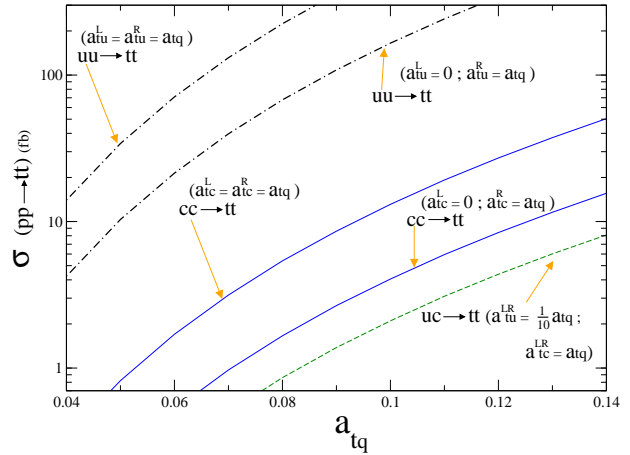}
\vspace*{0mm}
\caption{LHC production due to Ztc and Ztu FCNC couplings.
}
\label{prodz}
\end{figure}

We show in Fig.~\ref{prodz} the production cross section
for $pp\to tt$ for values of the $a^R_{tq}$ couplings below
0.15 ($X_{tq} \leq 0.4$). With an expected 100 fb$^{-1}$ of
luminosity there could be a few hundreds pairs of same sign
tops coming from this coupling.  Notice
that the scale of the plot Fig.~(\ref{prodz}) is logarithmic
to accomodate the $a_{tq}^4$ dependence of the cross section.  
Because of the larger parton luminosity from the $u$ quark
the process
$uu\to tt$ is more sensitive to the size of the FCNC couplings.
However, it is usually assumed that the top-charm transition
coupling should be the larger one (as it turns out with the
hierarchy of the CKM matrix elements).  We could have run
the values of $a^R_{tc}$ up to 0.3 but we have only considered
values up to 0.14 to prove that indeed the production rate
could be large enough.  In fact, in Ref.~\cite{slabospitsky}
the authors have used $a^R_{tu}$ of the order 0.4 and have
obtained a huge rate of a couple tens of pb for the LHC.

To bear in mind, for the $cc\to tt$ process there is also
the equivalent $\bar c\bar c\to \bar t\bar t$ like sign
anti-top production process, because the parton luminosity of
$c$ and $\bar c$ coincide.  According to Fig.~(\ref{prodz})
if the bounds of Ref.~\cite{extraquarks} apply for this
case the production from charm quarks could be not higher than
of order 1 fb.  As we shall discuss later a rate of only a few
fb's may be too small to ever be identified at the LHC.  We
should consider rates of order 10 fb and higher the ones to
look for to make the search of this signal feasible.
In this sense, the production of same sign top quark pairs
is not the best way to probe these FCNC couplings, in fact
the associated production of a single top with a neutral boson
like Z or $\gamma$ is considerably more sensitive.  However,
$tt$ production is an interesting signal that clearly
indicates the presence of significant FCNC.

To be consistent with the usual experimental limitations,
in this work we have imposed the following cuts on the
rapidity, transverse momentum and invariant mass of the
top quarks:
\ba
|y_t| &&\leq 2 \nonumber \\
Pt_t &&\geq 40  \rm{GeV} \nonumber \\
M_{tt} &&\geq 380 \rm{GeV} \nonumber
\ea
These cuts reduce the total rate in about $40\%$.

We have taken the factorization scale equal to the
invariant mass of the $tt$ pair $\mu = M_{tt}$.
The top quark mass we have used is $m_t = 175$GeV.
The Parton Distribution Function (PDF) used was the recent
CTEQ6M\cite{lai}.  Our calculations are at tree level,
obtained with the aid of the CompHEP program\cite{pukhov}.

Please notice that for the process $uc\to tt$ it is
understood that the other possibility $cu\to tt$
is already included in the figures shown.

\subsection{FCNC with the Z' boson}

\begin{figure}
\includegraphics[width=8.1cm,height=6cm]{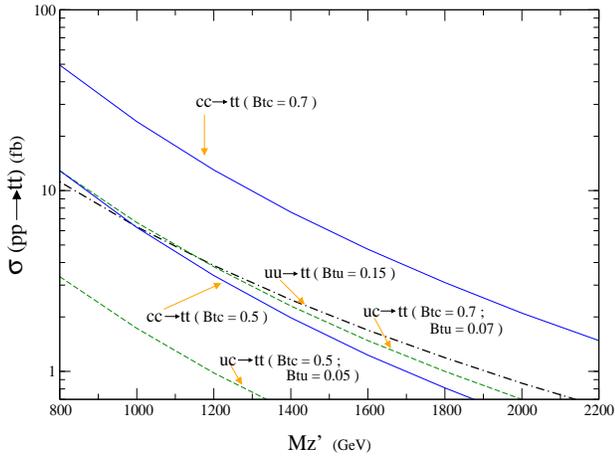}
\vspace*{0mm}
\caption{LHC production due to Z'tc and Z'tu FCNC couplings.
}
\label{prodzp}
\end{figure}

Now we want to consider a heavy Z' boson, such as the one
that appears in Topcolor assisted technicolor (TC2)
models\cite{chivukula}.  In these models the heavy Z' couples
strongly with the third family of quarks and may induce FCNC.
The coupling strength is given by
$g_2\, =\, e/c_W s_\phi c_\phi$ where $c_W$ is the cosine of
the SM $SU(2)_L \times U(1)_Y$ mixing angle and $\phi$ is
the mixing angle for the $Z_1$ and $Z_2$ neutral bosons
of the TC2 model.  As shown in Ref.~\cite{chivukula}
electroweak data requires the mass of Z' to be higher than
1-4 TeV for values of $\sin^2 \phi$ between 0.05 and 0.5
(this range of values implies the diagonal coupling strength
$g_2$ to be of order between 0.7 and 1.2).
We show in Fig.~(\ref{prodzp}) the production cross section
of $pp \to tt$ for different values of $M_{Z'}$.
We have chosen high values of the coupling $B_{tc}$
(between 0.5 and 0.7) but we have also assumed only the
right handed current term $B^R_{tc}$ as different from zero.
Given the assumption that the cross section
must be at least of order 10 fb to become detectable we
can see that $B^R_{tc}$ must be higher than 0.5 to yield
this production rate.  Again, other processes where the
anomalous coupling appears only once in the Feynman diagram
are more sensitive.  For instance, from Ref.~\cite{yue}
a loop level induced rare top decay $t\to cV$
($V=\gamma, Z$ or gluon) would require a smaller coupling
size $B_{tc} = g_1 K_{tc} \sim 0.2$ to become of order $10^{-5}$
(for gluon) or $10^{-6}$ (for $\gamma$) for its branching ratio,
which is already at expected detectable levels at the LHC.
However, the rare top quark decay would be an indirect way
to test the existence of Z'.

Finding a Z', or any heavy resonance, that couples preferentially
to the third family is possible via $\tau$ lepton pair
production\cite{mrenna}, but it may not be possible at all at any
future hadron collider (including the LHC) searching in the
$t\bar t$ mode\cite{valencia}.  This is because of the
overwhelming QCD production of $t\bar t$, which in contrast makes
no background for a $tt$ signal.  Therefore, either the single
top production mode or the same sign $tt$ pair mode could be the
way to test this kind of physics at the LHC.

\subsection{Scalar FCNC}

\begin{figure}
\includegraphics[width=8.1cm,height=6cm]{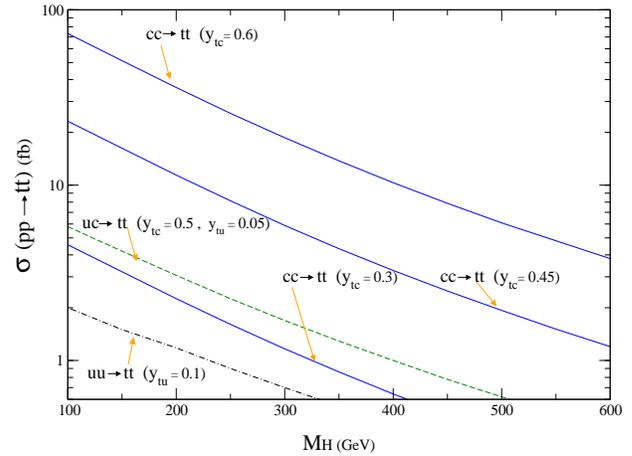}
\vspace*{0mm}
\caption{LHC production due to Htc and Htu FCNC couplings.
}
\label{smh}
\end{figure}

Next we want to consider a FCNC scalar coupling of the
type $Htq$\cite{hill}.
This coupling was also studied in Ref.~\cite{branco, tcatlhc}.
It is not expected that the $y^L_{tc}$ coupling could be
very large, maybe of order $\lambda^2 m_t/f_{\pi}$ with
$\lambda = 0.22$ in relation
to the CKM mixing parameter (the Cabibbo angle).  Also,
$y^L_{tu}$ would be less than $\lambda^3 m_t/f_{\pi}$
\cite{kanemura, burdman}.
However, there is no reason to believe the right handed
$y^R_{tq}$ coupling couldn't be much larger.  In fact,
the dynamical Top-color (TopC) model generally predicts
$0.2\leq y^R_{tc}\leq 0.7$\cite{kanemura, burdman}.
For this size of the FCNC coupling the production of top
pairs can be significant, see Fig.~(\ref{smh}).

If the scalar Higgs mass is not very high,
between 100 and 200 GeV, a coupling somewhat higher
than $y^R_{tc}=0.3$ could give enough rate to
be observed at the LHC.

\subsubsection{Scalar FCNC production at a VLHC}

\begin{figure}
\vspace*{10mm}
\includegraphics[width=8.1cm,height=6cm]{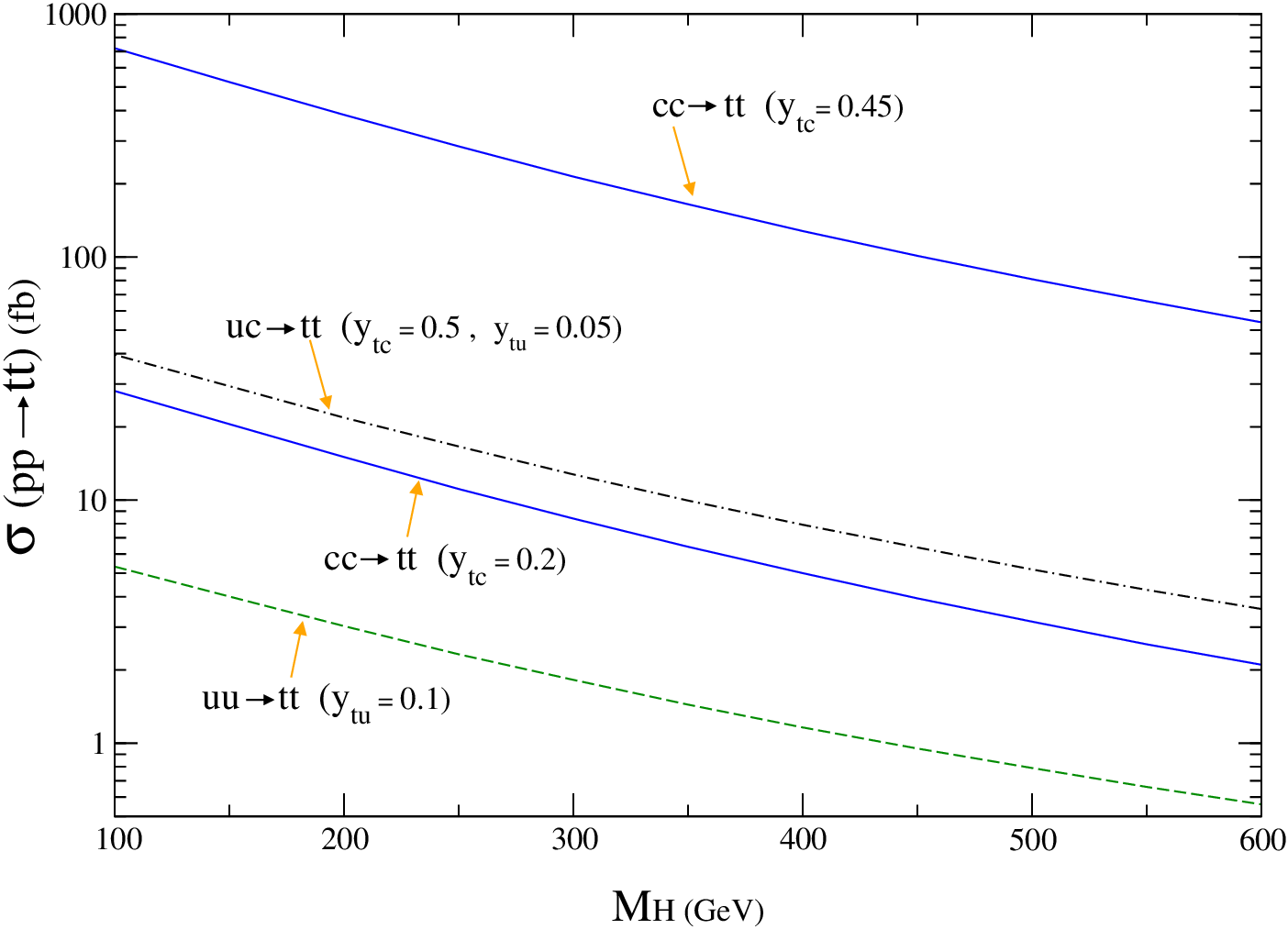}
\vspace*{0mm}
\caption{VLHC production due to Htc and Htu FCNC couplings.
The CM energy of the collider is taken as $\sqrt{s} = 100$TeV.
}
\label{vsmh}
\end{figure}

At this point it is interesting to compare with what
would be expected at a VLHC machine with a center of
mass (CM) energy of $\sqrt{s} = 100$TeV 
for this same scalar coupling.  In Figure~(\ref{vsmh})
we show the much higher rates for even smaller values
of the coupling size and for a heavier scalar boson.

\subsection{Dependence on PDF set and factorization scale.}

\begin{figure}
\includegraphics[width=8.1cm,height=6cm]{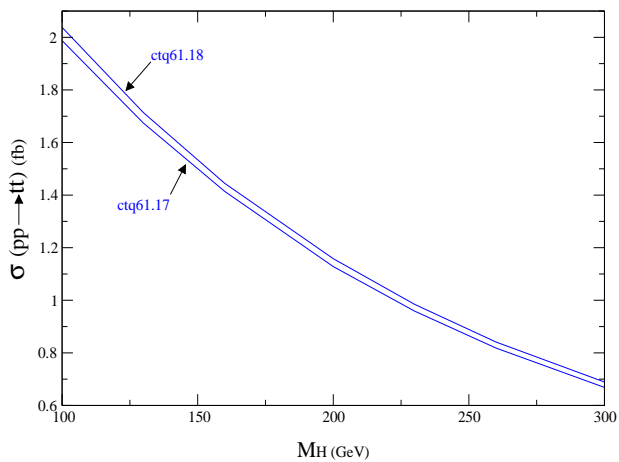}
\vspace*{0mm}
\caption{LHC production of tt via $uu\to tt$,
varying the PDF used.  The factorization scale $\mu$ is taken
constant at 380 GeV. 
}
\label{pdfuu}
\end{figure}

Throughout this work we have used CTEQ6M.  We have checked
the uncertainty due to the PDF by computing the production 
of tt pairs with all the 41 sets of ctq6l PDFs
provided by the CTEQ group\cite{lai}.  The variation
amounts to only about 10$\%$, we show our results in
Figure~(\ref{pdfuu}).
We have only shown the computations that are farther apart
from each other, i.e. the results of any other PDF set will
fall between the curves shown.  Figure~(\ref{pdfuu}) applies
to the $uu\to tt$ process; a similar graph would apply to
the $cc\to tt$ process, except that the highest (lowest)
curve would come from ctq6l.09 (ctq6l.10) instead of the
ctq6l.18 (ctq6l.17) PDF set.

To compare different values of the factorization scale we
have tried some fixed scale values, it turns out that the
fixed value of $380$ GeV yields almost the same cross
section as the running  $\mu = M_{tt}$ value.  Also,
there is some mild dependence on the (fixed) factorization
scale, we have found about less than $20\%$ variation
going from $\mu = 190$GeV to $\mu = 760$GeV.

\section{Identification of tt pairs at LHC}

Compared with the number of $t\bar t$ pairs that
are expected every year at the LHC, of the order of
10 million, our $tt$ signal is overwhelmed by many
orders of magnitude.  There are 3 modes of detecting
a pair of top quarks depending on the decay products
of the $W$ bosons: dilepton, lepton plus jets and all
jets mode.  Separating these signals can only happen via
the dilepton mode, in which the two same sign tops
produce two same sign leptons.
Unfortunately, this mode is associated with a small
branching ratio, approximately $5\%$. Considering a
$50\%$ b-tagging efficiency, this mode will let us
observe only a small $2\%$ (or less) fraction of all
the top quark pairs produced\cite{abe,topd0}.
Our assumption here is that it will
be possible to identify the sign of top quark pairs
via the dilepton mode with an overall efficiency of
order $1\%$ at the LHC.  Then, we can expect to observe
from a few to up to several tens of same sign top events
every year for the parameters used in this paper.

\subsection{SM background for $tt$ production.}

\begin{figure}
\includegraphics[width=5.1cm,height=3cm]{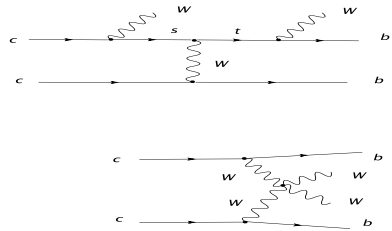}
\vspace*{0mm}
\caption{The two most important diagrams for the SM
production of $bbW^+W^+$ which is the background signal
for $tt$ production.   
}
\label{background}
\end{figure}

The SM background we have considered here is production
of $pp\to bbW^+W^+$.
In the SM there are 45 diagrams for the $uu\to bbW^+W^+$
process (49 for $cc\to bbW^+W^+$), of which two give the
most important contributions, see Figure (\ref{background}).
By taking only one diagram at a time we have found that the
greatest contribution is less than of order
$6\times 10^{-6}$ fb. So, we can safely conclude that the
total contribution must be well below the $10^{-4}$ fb level.
This is a negligible background.

\section{Conclusion}

In this work we have shown that FCNC couplings of the top
quark with either the SM gauge bosons or new bosons like the
ones that appear in Top color assisted technicolor theories,
could give rise to significant production of same sign top
quark pairs at the LHC.  Such a signal could be searched for
in the dilepton mode.  We have found that for FCNC top quark
couplings, well within the present experimental constraints,
we could have production cross sections of order several tens
or a few hundred femtobarns at the LHC.   In terms of
sensitivity, the cross section is proportional to the fourth
power of the FCNC coupling.  In particular, a coupling size
of order 0.05 (0.1) for tuZ (tcZ) would give rise to
a few tt pairs detected in the dilepton mode at the LHC.
In addition, in the case of a heavy resonance like Z' or
a scalar H, a coupling size of 0.15 (0.5) for tuZ'
(tcZ') along with a Z' mass between 1 and 1.6 TeV would
be required for the same yield. A coupling size higher than
0.1 (0.3) for tuH (tcH) along with a 100-300 GeV mass
range in the case of the scalar FCNC;  in comparison, for
even slightly smaller couplings a twice as large mass range
of 100-600 GeV of the scalar boson could be probed in a
VLHC with CM energy $\sqrt{s}=100$TeV.

\vspace{2cm}

\begin{figure}
\includegraphics[width=8.1cm,height=6cm]{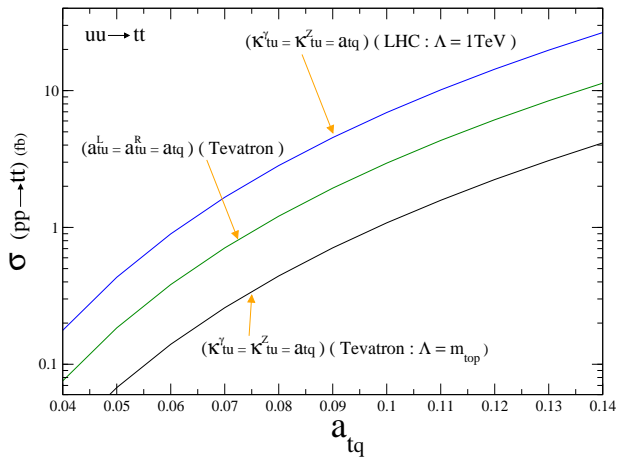}
\vspace*{0mm}
\caption{Tevatron and LHC production via $uu\to tt$ from
dimension 5 tuZ, tu$\gamma$ couplings.  The factorization
scale used is $\mu=M_{tt}$ the invariant mass of $tt$. The
couplings $\kappa^\gamma$ and $\kappa^Z$ are defined in
Ref.~\cite{kidonakis}.  No cuts on the $tt$ signal are
applied in this case.  The production rate from the dimension
4 couplings used by us is shown for comparison.
}
\label{kido}
\end{figure}

Upon the completion of this article, another work on the
production of top quarks via the FCNC anomalous dimension 5
tq$\gamma$ and tqZ couplings has appeared.  They have
computed the cross section for $p\bar p\to tt$ at the
Tevatron with NLO-NLL and NNLO-NLL corrections.  They
obtain a cross section of order a few femtobarns at the
Born level with a $25\%$ increase due to the higher order
corrections\cite{kidonakis}.  To compare with their analysis
we show in Figure~(\ref{kido}) the production of $tt$ at
both the Tevatron and the LHC from either, the combination
of dimension 5 tu$\gamma$ and tuZ couplings used by them
($\kappa^\gamma = \kappa^Z$) or the dimension 4 tuZ used
by us ($a_{tu}$).

Instead of the fixed value for the factorization
scale $\mu = m_t = 175$GeV, we have used $\mu = M_{tt}$.
That is why we have a somewhat lower (about $20\%$ less)
value for the Born level cross section.  Otherwise, if we
use $\mu =m_t$ we obtain the same rates.  To be consistent
with their calculations, we have applied no cuts in this figure.
The dimension 5 operator requires a mass scale factor $1/\Lambda$
for which they have used a rather small value $\Lambda = m_t$.
If we use the same scale $\Lambda = m_t$ for the LHC we
would obtain a huge rate of order $10^4$fb, and we could as
well be violating unitarity constraints.
For these operators, we think a higher mass scale
$\Lambda = 1$TeV is more realistic; we have used it for the
LHC production in Figure~(\ref{kido}).
Notice that the dimension 4 tuZ coupling
could yield even higher production rates than the dim 5
coupling.  However, due to the much smaller range of
luminosity of the Tevatron, as compared with the LHC, these
rates may be too low to be detectable.

\vspace*{3mm}
\noindent
{\bf Acknowledgments}~~~
We want to thank C.-P. Yuan for numerous suggestions for
this work.  We also thank Conacyt for support.

\vspace*{-4mm}



\begin{thebibliography}{1}


\bibitem{konisberg}
Chakraborty, D., Konigsberg, J. and Rainwater, D.,
hep-ph/0303092, to appear in the
Annual Review of Particle Science {\bf 53};
M. Beneke, I. Efthymiopoulos, M.L. Mangano, J. Womersley,
et al., {\it Top Quark Physics},
1999 CERN Workshop on the SM Physics (and more) at the LHC,
hep-ph/0003033.



\bibitem{raretopdec}
T. Han, R.D. Peccei and X. Zhang,
Nucl. Phys. {\bf B454}, 527 (1995);
T. Tait and C.-P. Yuan,
Phys. Rev. D{\bf 55}, 7300 (1997);
J.L. Diaz-Cruz, M.A. P\'erez, G. Tavares-Velasco and
J.J. Toscano, Phys. Rev. D{\bf 41}, 891 (1990); ibid,
Phys. Rev. D{\bf 60}, 115014 (1999);
E. Jenkins, Phys. Rev. D{\bf 56}, (1997) 458;
B. Mele, hpe-ph/0003064;
C. Yue, G. Liu and Q. Xu,
Phys. Lett. B{\bf 508}, 290 (2001);
T. Han, M. Hosch, K. Whisnant, B.L. Young and X. Zhang,
Phys. Rev. D{\bf 58}, 073008 (1998);
R. Diaz, R. Martinez, J. Alexis-Rodriguez, hep-ph/0103307.



\bibitem{multilepton}
F. del Aguila and J. Aguilar-Saavedra,
Nucl. Phys. {\bf B576}, 56 (2000).

\bibitem{branco}
J. A. Aguilar-Saavedra and G. C. Branco,
Phys. Lett. B{\bf 495}, 347 (2000).


\bibitem{eetop}
J.A. Hewett, Int. J. Mod. Phys. A13 (1998) 2389;
D. Atwood, L. Reina and A. Soni,
Phys. Rev. D{\bf 53}, 1199 (1996);
S. Bar-Shalom and J. Wudka,
Phys. Rev. D{\bf 60}, 094016 (1999);
C.-S. Li, X.-M. Zhang and S.-H. Zhu,
Phys. Rev. D{\bf 60}, 077702 (1999);
G. Couture, Phys. Rev. D{\bf 62}, 097503 (2000);
C.-X. Yue, Y.-B. Dai, Q.-J. Xu and G.-L. Liu,
Phys. Lett. B{\bf 525}, 301 (2002);
C.-S. Huang, X.-H. Wu and S.-H Zhu,
Phys. Lett. B{\bf 452}, 143 (1999);


\bibitem{hill}
C.T. Hill, Phys. Lett. B{\bf 345}, 483 (1995).

\bibitem{theories}
R.S. Chivukula, E.H. Simmons and J. Terning,
Phys. Lett. B{\bf 331}, 383 (1984);
D.J. Muller and S. Nandi,
Phys. Lett. B{\bf 383}, 345 (1996);
E. Malkawi, T. Tait and C.-P. Yuan,
Phys. Lett. B{\bf 385}, 304 (1996);
K. Lane and E. Eichten,
Phys. Lett. B{\bf 433}, 96 (1998);
R.S. Chivukula, B.A. Dobrescu, H. Georgi and C.T. Hill,
Phys. Rev. D{\bf 59}, 075003 (1999);
H. Georgi and A.K. Grant,
Phys. Rev. D{\bf 63}, 015001 (2001).


\bibitem{slabospitsky}
Y.P. Gouz and S.R. Slabospitsky,
Phys. Lett. B{\bf 457}, 177 (1999).


\bibitem{tcgluon}
T. Han, K. Whisnant, B.L. Young and X. Zhang,
Phys. Lett. B{\bf 385}, 311 (1996).
E. Malkawi and Tim Tait,
Phys. Rev. D{\bf 54}, 5758 (1996).
H. Fritzsch, Phys. Lett.{\bf 224}, 423 (1989).


\bibitem{wwagner}
W. Wagner, by the CDF collaboration, {\it Top Quark Physics
with CDF}, Presented at 14th Topical Conference on Hadron Collider
Physics (HCP2002) Karlsruhe, Germany, Sep/29 - Oct/4, 2002.
FERMILAB-Conf-02/317-E.


\bibitem{abe}
F. Abe, et. al.
Phys. Rev. Lett. {\bf 80}, 2525 (1998);
See also S. Cabrera, hep-ex/0305066 for a recent report of
the D0 and CDF collaborations.

\bibitem{extraquarks}
F. del Aguila, J. A. Aguilar-Saavedra and J. Miquel,
Phys. Rev. Lett. {\bf 82}, 1628 (1999).


\bibitem{lai} J. Pumplin, D.R. Stump, J. Huston, H.L. Lai,
P. Nadolsky and W.K. Tung.  JHEP 0207 (2002) 012.  Also see,
hep-ph/0307022.

\bibitem{pukhov} A. Pukhov, E. Boos, M. Dubinin, V. Edneral,
V. Ilyin, D. Kovalenko, A. Kryukov, V. Savrin, S. Shichanin,
A. Semenov,  CompHEP -a package for evaluation of Feynman
diagrams and integration over multiparticle phase space,
hep-ph/9908288.


\bibitem{chivukula}
R.S. Chivukula and E.H. Simmons,
Phys. Rev. D{\bf 66}, 015006 (2002).


\bibitem{yue}
C.-X. Yue, H. Zong and G.-L. Liu, hep-ph/0309255.


\bibitem{mrenna}
K. Lynch, S. Mrenna, M. Narain and E.H. Simmons,
Phys. Rev. D{\bf 63}, 035006 (2001).

\bibitem{valencia}
T. Han, D. Rainwater and G. Valencia,
Phys. Rev. D{\bf 68}, 015003 (2003).


\bibitem{tcatlhc}
J. Cao, Z. Xiong and J.M. Yang,
Phys. Rev. D{\bf 67}, 071701 (2003).


\bibitem{kanemura}
H.-J. He, S. Kanemura and C.-P. Yuan,
Phys. Rev. Lett. {\bf 89}, 101803 (2002);
H.-J. He, and C.-P. Yuan,
Phys. Rev. Lett. {\bf 83}, 28 (1999).

\bibitem{burdman}
G. Burdman,
Phys. Rev. Lett. {\bf 83}, 2888 (1999);


\bibitem{topd0} The D0 Collaboration,
Phys. Rev. D{\bf 67}, 012004 (2003).


\bibitem{kidonakis}
N. Kidonakis and A. Belyaev, hep-ph/0310299.



\end{thebibliography}
\end{document}